\documentclass[]{aastex631}

\begin{document}

\title{Investigating Temporal and Spatial Variation of Jupiter’s Atmosphere with Radio Observations
}

\author[0009-0001-2730-1263]{Joanna Hardesty}
\affiliation{Department of Astronomy \\
501 Campbell Hall, University of California, Berkeley, CA 94720, USA}

\author[0000-0002-6293-1797]{Chris Moeckel}
\affiliation{Department of Earth and Planetary Science \\
307 McCone Hall, University of California, Berkeley, CA 94720, USA}

\author[0000-0002-4278-3168]{Imke de Pater}
\affiliation{Department of Astronomy \\
501 Campbell Hall, University of California, Berkeley, CA 94720, USA}

\begin{abstract}
\noindent  We study the spatial and temporal variability in Jupiter’s atmosphere by comparing longitude-resolved brightness temperature maps from the Very Large Array (VLA) radio observatory and NASA’s Juno spacecraft Microwave Radiometer (MWR) taken between 2013 and 2018. Spatial variations in brightness temperature, as observed at radio wavelengths, indicate dynamics in the atmosphere as they trace spatial fluctuations in radio-absorbing trace gases or physical temperature. We use four distinct frequency bands, probing the atmosphere from the water cloud region at the lowest frequency to the pressures above the ammonia cloud deck at the highest frequency. We visualize the brightness temperature anomalies and trace dynamics by analyzing the shapes of brightness temperature anomaly distributions as a function of frequency in Jupiter's North Equatorial Belt (NEB), Equatorial Zone (EZ), and South Equatorial Belt (SEB). The NEB has the greatest brightness temperature variability at all frequencies, indicating that more extreme processes are occurring there than in the SEB and EZ. In general, we find that the atmosphere at 5 and 22 GHz has the least variability of the frequencies considered, while observations at 10 and 15 GHz have the greatest variability. When comparing the size of the features corresponding to the anomalies, we find evidence for small-scale events primarily at the depths probed by the 10 and 15 GHz observations. In contrast, we find larger-scale structures deeper (5 GHz) and higher (22 GHz) in the atmosphere.
\end{abstract}


\section{Introduction} \label{sec:intro}
There are many examples of processes on giant planets that can alter the atmosphere and affect the temperature and distribution of trace gases. In this paper, we use spatial variations in brightness temperature as a proxy for dynamics in Jupiter's atmosphere. At radio wavelengths, the effect of clouds is small, so we can probe the atmosphere within and below cloud layers \citep{dePater2019_VLA}. At these wavelengths, we measure a planet's brightness temperature, the temperature an equivalent blackbody would have to match the observed flux density.
Brightness temperature depends on the temperature-pressure profile in the atmosphere (for Jupiter, usually assumed to be adiabatic at pressures P $>$ 0.8 bar) and the abundance of radio-absorbing gases (on Jupiter, at $\sim$1--25 GHz, this is primarily NH$_3$ gas, but H$_2$S and H$_2$O also absorb at radio wavelengths) \citep{dePater1986Jzsa,depater2023RemS}. Since the brightness temperature is inversely proportional to the amount of trace gases in the atmosphere, enhancements and depletions of these gases can be studied from the brightness temperature profile and used as indicators of dynamics in the atmosphere. Negative or positive brightness temperature anomalies can indicate upwelling or downwelling of radio-absorbing trace gases in the atmosphere when the background tracer gas has a vertical concentration gradient \citep{Gierasch1986, dePater2016}. Alternatively, anomalies can point to physical temperature variations in the atmosphere. Either interpretation, physical temperature fluctuation or abundance variation, indicates the dynamics we are interested in in this paper. \par

Variations in brightness temperature with latitude can be seen on all four giant planets, generally interpreted as latitudinal variations in the radio-opacity sources mentioned above \citep{depater2023RemS}. These variations have been attributed to dynamics, though we still lack a full explanation. One example of large-scale variations is Saturn, where giant storms cause vertical transport of ammonia gas and show long-lasting signatures \citep{LiCheng2023Ldeo}. These storms can disrupt convection patterns and potentially explain the observed depletion of ammonia in the upper few bars of the atmosphere through condensation, precipitation, subsiding dry air, and potential horizontal motions \citep{ShowmanAdamP.2005DIoJ}. \citet{GuillotTristan2020mushballs} hypothesized the formation and subsequent precipitation of ammonia-rich hail or ``mushballs” to deplete condensible gases in Jupiter's atmosphere to well below the level of the water cloud base near 6 bar. \par

We aim to characterize brightness temperature probability distributions to further inform our understanding of atmospheric dynamics on Jupiter. We will study temporal and spatial variations by comparing brightness temperature measurements between 2013 - 2018. We use data from two instruments: the Microwave Radiometer (MWR) on board the NASA Juno spacecraft \citep{Janssen2017} and the Very Large Array (VLA) radio observatory. We use three distinct frequencies, where both instruments overlap in coverage and an additional VLA frequency where there are no MWR measurements, to map out anomalies in the equatorial region of Jupiter, covering latitudes from the South Equatorial Belt (SEB) to the North Equatorial Belt (NEB). 

In Section \ref{sec:data}, we present the MWR and VLA data used in the analysis. In Section \ref{sec:methods}, we explain the process for determining the zone and belt boundaries, the effect of limb darkening on our analysis, weighting functions, and the resolution of our measurements. In Section \ref{sec:results}, we describe the results of our analysis, highlighting the differences between the tropical belts and the Equatorial Zone (EZ) and visualize the scale of the observed extreme features. Finally, Section \ref{sec:conclusions} summarizes our conclusions and recommendations for future work.

\section{Data} \label{sec:data} 
\subsection{Juno MWR} \label{subsec:juno}

MWR on board the Juno spacecraft is equipped with six radiometers that measure radiation between 0.6 GHz and 22 GHz centered around six distinct frequencies, referred to as Channels \citep{Janssen2017}. We make use of Channel 4 (C4, 5.2 GHz), Channel 5 (C5, 10.0 GHz), and Channel 6 (C6, 21.9 GHz) on MWR, frequencies which overlap with observations taken by the VLA. At radio wavelengths, we measure the radiant flux integrated over a column of atmosphere, whose depth is determined by the opacities of trace gases in the atmosphere. The spatial resolution on the planet depends on the wavelength, the size of the antenna (or spacings between antennas for an interferometer), and the distance of the instrument to the target. The received flux density is then converted into units of equivalent brightness temperature. The VLA requires long integration times to build up sufficient signal-to-noise while Jupiter rotates. Juno's approach differs, with the spacecraft spinning perpendicular to the spin axis of Jupiter, which results in multiple measurements per latitude at varying geometries but confined to a narrow longitudinal swath of the planet. The MWR brightness temperatures were obtained through an iterative approach that deconvolves the measurements based on overlapping observations \citep{Moeckel2024}, similar to the approach developed by \citet{zhang2020}.

\begin{figure}[h!]
\epsscale{1.0}
\plotone{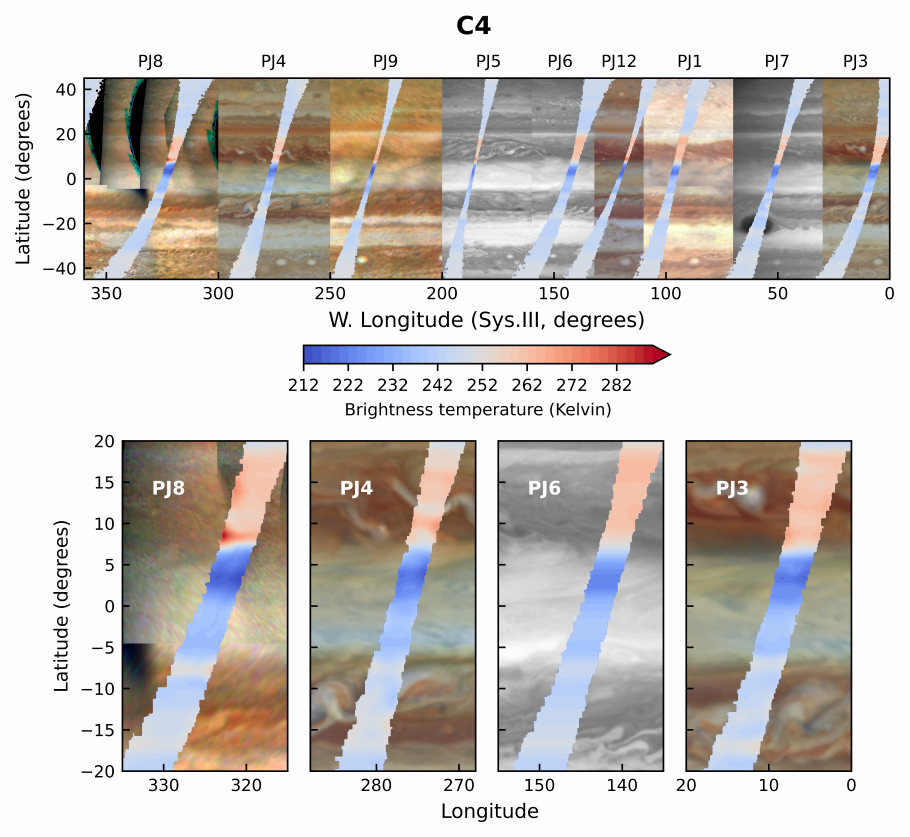}
\caption{Juno MWR C4 brightness temperature measurements for each perijove superimposed on image slices of Jupiter, taken close to the time that the flyby occurred and plotted between +/- 45{\textdegree} planetographic latitude. Further information about the visible image slices can be found in Table \ref{tab:Obs}. All MWR brightness temperature values correspond to the same color bar at the center of the image. The HST images are all from project ID GO-14661 \citep{WongMichaelH.2020HUIo} and were obtained from the MAST archive at\dataset[doi: 10.17909/T94T1H]{https://doi.org/10.17909/T94T1H}. Credit for JunoCam maps (PJs 1, 8, and 9): NASA / JPL / SwRI / MSSS / Gerald Eichstädt / John Rogers.}
\label{fig:fig1_C4}
\end{figure}

\begin{figure}[h!]
\epsscale{1.0}
\plotone{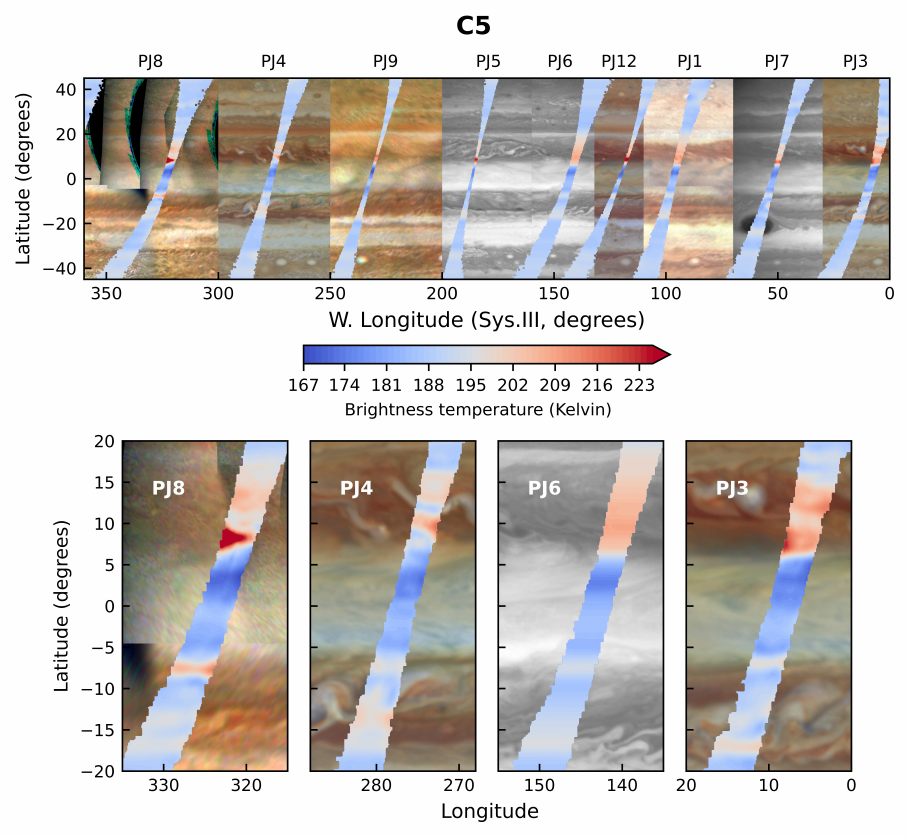}
\caption{Similar to Figure~\ref{fig:fig1_C4}, but for Juno MWR C5.}
\label{fig:fig1_C5}
\end{figure}

\begin{figure}[h!]
\epsscale{1.0}
\plotone{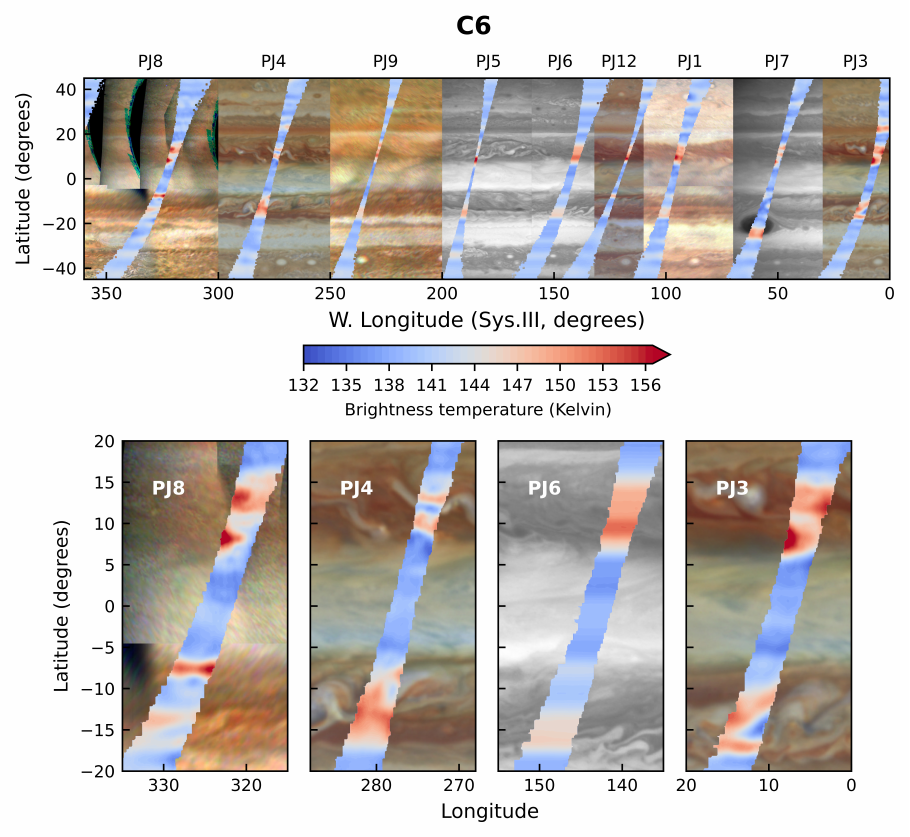}
\caption{Similar to Figure~\ref{fig:fig1_C4}, but for Juno MWR C6.}
\label{fig:fig1_C6}
\end{figure}

Figures \ref{fig:fig1_C4}, \ref{fig:fig1_C5}, and \ref{fig:fig1_C6} show the longitude-resolved nadir brightness temperature measurements of the first 12 perijoves (PJs) for MWR C4, C5, and C6, respectively, between +/- 45{\textdegree} planetographic latitude. The HST images in the three figures are from the Mikulski Archive for Space Telescopes (MAST)\footnote{The HST images from MAST can be found at\dataset[doi: 10.17909/T94T1H]{https://doi.org/10.17909/T94T1H}}. These measurements represent the first time Juno MWR measurements have been combined and displayed to highlight the longitudinal brightness temperature structure that MWR can capture. We plot the \textbf{brightness} temperature distribution onto concurrent, visible observations taken with the Hubble Space Telescope (HST) and JunoCam as shown in Table \ref{tab:Obs}. The figure clearly shows \textbf{brightness} temperature trends between the zones and belts, with the EZ being cooler than the NEB and SEB on average, including smaller \textbf{brightness} temperature variations in longitude. \par

In PJ4, a storm at $\sim$12{\textdegree}N latitude appears in the visible images and is captured well by the three MWR channels. Another storm is captured near $\sim$12{\textdegree}S and discussed in detail by \cite{Moeckel2024}. Longitude-resolved hot spots are seen at northern latitudes in PJ3 and PJ8, particularly prominent in C5 and C6. PJ3 and PJ8 also have interesting structures at southern latitudes (PJ3 traverses the wake of the Great Red Spot).

\begin{table}[h]
\caption{Instruments used for Jupiter images in Figures~\ref{fig:fig1_C4}, \ref{fig:fig1_C5}, and \ref{fig:fig1_C6}. \label{tab:Obs}}
\hspace{40mm}
\begin{tabular}{ccccc}
\hline
\hline
PJ \# & Observatory & PJ date        & PJ time (UTC) \\
\hline
1     & JunoCam     &   2016/08/27   & 12:50\\
3     & HST         &   2016/12/11   & 17:03 \\
4     & HST         &   2017/02/02   & 12:57\\
5     & HST         &   2017/03/27   & 08:51\\
6     & HST         &   2017/05/19   & 06:00\\
7     & HST         &   2017/07/11   & 01:54\\
8     & JunoCam     &   2017/09/01   & 21:48\\
9     & JunoCam     &   2017/10/24   & 17:42\\
12    & HST         &   2018/04/01   & 17:56\\
\hline
\end{tabular}
\break
\break
\raggedright

Each HST and JunoCam image listed in this table was cropped and combined to make Figures~\ref{fig:fig1_C4}, \ref{fig:fig1_C5}, and \ref{fig:fig1_C6}. All dates are formatted as year/month/day. The HST images were taken during the PJ indicated. For PJ3 and PJ4, the HST images were taken using a 395 nm filter, while PJ5, PJ6, and PJ7 utilized a combination of 395 nm, 502 nm, and 631 nm filters.
\end{table}

\subsection{VLA} \label{subsec:vla}

\begin{table}[h]
\footnotesize
\caption{VLA parameters. \label{tab:VLA}}
\centering
\begin{tabular}{cccccccccc}
\hline
\hline

 Band & Freq. Range  & Bandwidth & $T_b$ (peak)& $p$  & Date            & Config.& Resolution & Major beam- & Minor beam- \\
      &  (GHz)       &    (GHz)  &      (K)    &      & (year/month/day)&        &   (km)     &  width (mas)& width (mas) \\
\hline
C    & 4 - 8         & 4         & 240         & 0.16 & 2014/05/04      & A      & 2400       & 588         & 588   \\
X    & 8 - 12        & 4         & 186         & 0.16 & 2014/02/09      & B      & 3300       & 1036        & 693   \\
X    & 8 - 10        & 2         & 181         & 0.16 & 2016/12/16      & A      & 1100       & 252         & 165   \\
Ku   & 12 - 18       & 6         & 154         & 0.08 & 2013/12/23      & B      & 1800       & 579         & 359   \\
K    & 18 - 26       & 8         & 136         & 0.065& 2014/12/27      & C      & 4000       & 1217        & 766   \\
\hline
\end{tabular}
\break
\break
\small
\raggedright
Frequency band, frequency range, bandwidth, peak brightness temperature, limb-darkening coefficient, observation date, array configuration, resolution, major axis beamwidth, and minor axis beamwidth of the VLA data used for analysis. The 2013 - 2014 VLA data in this paper were already published in \citet{dePater2016} and \citet{dePater2019_VLA} and the 2016 VLA data was published in \citet{Moeckel2023}. 
\end{table}
 
In addition to the VLA observations acquired in 2013 - 2014 \citep{dePater2016,dePater2019_VLA}, we use VLA data obtained in 2016, which were first shown by \citet{Moeckel2023}. These data were taken in conjunction with Juno's PJ3 (Dec. 16th, 2016), when the VLA was in its most extended A configuration, making them the highest-resolution radio observations of Jupiter to date. The radio emissions were integrated over 9 hours, nearly sampling a full planet rotation. The observations were run through the standard pipeline in CASA (Common Astronomy Software Applications, v.5.1.0) with 3C286 as the flux density calibrator and J1246-0730 as the phase calibrator \citep{CASA_2022}. Visibilities from noisy antennas and baselines were deleted. After averaging the visibilities in frequency (8 channels, 16MHz) and time (10s), the phases of the observations were self-calibrated on a model atmosphere (disk-averaged brightness temperature of $T_{b}$, a limb-darkening coefficient of $p$ following the $cos(\theta)^p$ taken from \citet{dePater2019_VLA}, values for $T_b$ and $p$ shown in Table~\ref{tab:VLA}). Due to the fact that there were no short baselines in the A configuration, the observations resolve the majority of the planet's disc, requiring the inclusion of the typical zone-belt structure for the self-calibration process. We added the zonal radio-residuals \citep{dePater2016} to our model and performed a single iteration of self-calibration. We then use the deprojection algorithm developed by \cite{Sault2004} to account for the planetary rotation to obtain longitude-resolved radio maps. 

\begin{figure}[h!]
    \centering
    \epsscale{1.2} 
    \plotone{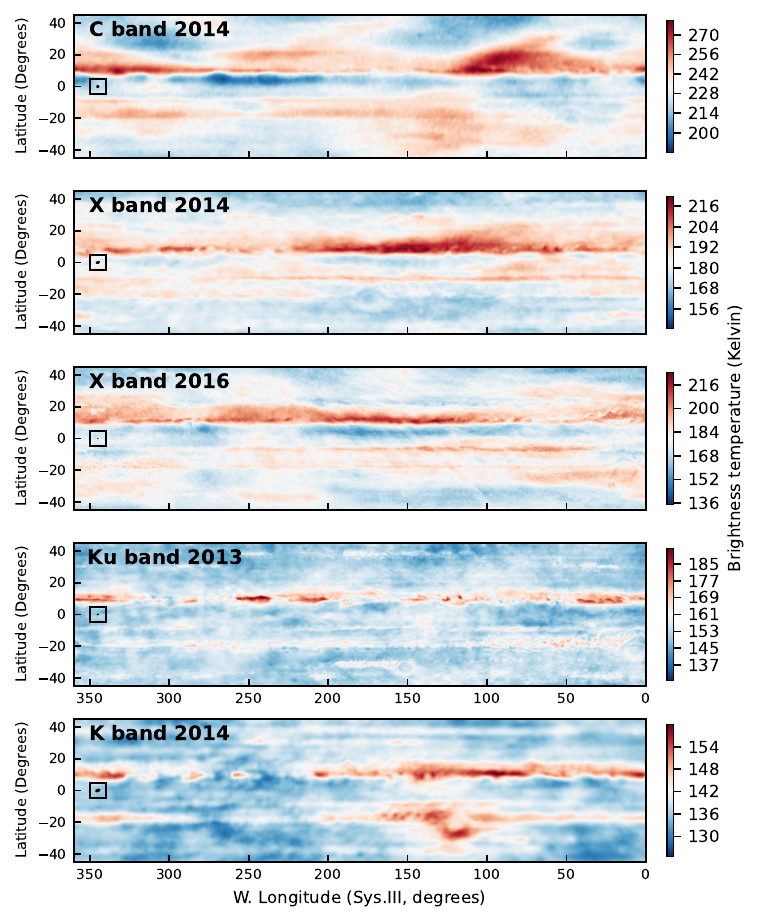}
    \caption{VLA maps of Jupiter with C, X, Ku, and K bands taken in 2013 - 2014 from \citet{dePater2016,dePater2019_VLA} and additional X band data from 2016 from \citet{Moeckel2023} as labeled. All maps are shown between +/- 45{\textdegree} planetographic latitude and all had a limb-darkened disk added back in to show the absolute brightness temperature distribution. The color bar represents the brightness temperature of the maps in units of Kelvin. The resolution ellipse for each map is shown at the equator in black within a black box. We refer the reader to \citet{dePater2016} to compare with semi-concurrent visible images.}
    \label{fig:VLA maps}
\end{figure}

\section{Methodology} \label{sec:methods}
\subsection{Latitude Boundaries} \label{subsec:zwp}
Jupiter’s atmosphere is divided into visibly dark “belts," which have cyclonic wind shear, and lighter “zones," which have anticyclonic wind shear. We expect to observe distinct atmospheric processes in each region of the planet due to differences in gas abundances, so we split our analysis accordingly. We use the peaks of the zonal wind profile, the locations at which the zonal wind profile reverses, as a proxy for the boundaries between the zones and the belts. We applied SciPy's Savitzky–Golay filter \citep{2020SciPy-NMeth} to the raw zonal wind profile from \citet{Tollefson2017} to smooth the data such that one value could be obtained for each zone or belt boundary\footnote{Zonal wind profile can be found at: \url{https://archive.stsci.edu/hlsp/wfcj}}. For the NEB we found the boundary to be 6{\textdegree} to 17{\textdegree} degrees latitude, the EZ from -7{\textdegree} to 6{\textdegree}, and the SEB from -20{\textdegree} to -7{\textdegree}. The results agree to within 1{\textdegree} with the zone and belt boundaries given by \citet{fletcher2017_CoAitJA}, confirming the remarkable stability of the zone-belt structure as derived from zonal wind profiles.

\subsection{Limb darkening} \label{subsec:limbdark}
The MWR has variable orbital geometry throughout each PJ, while the VLA observes Jupiter at a constant sub-observer latitude and distance from Earth's surface. One of the two instruments' measurements must be adjusted to directly compare VLA and MWR observations. While the original maps published by \citet{dePater2019_VLA} showed the radio-residuals to bring out the fine-scale structure, we added back the original limb-darkened disk to obtain absolute brightness temperature maps for the VLA (see Figure \ref{fig:VLA maps}), as they did before analyzing their data.

Instead of correcting the VLA observations to nadir brightness observations, we transformed the Juno observations to the VLA's viewing geometry. Juno measures limb darkening and nadir brightness temperature \citep{Janssen2017}. We can use MWR's limb darkening measurement, $p$, to convert the observed nadir brightness temperature, $T_{nadir}$, to an Earth-based geometry using $T_{b} = T_{nadir} \ cos(\theta)^p$ \citep{Moeckel2023}. This correction will decrease the extremes in the MWR measurements, but limb darkening is a minor effect for the frequencies and latitudes considered; the greatest difference before and after the correction was 1.1\%.

\subsection{Weighting functions}
\begin{figure}[]
    \epsscale{1.15} 
    \plotone{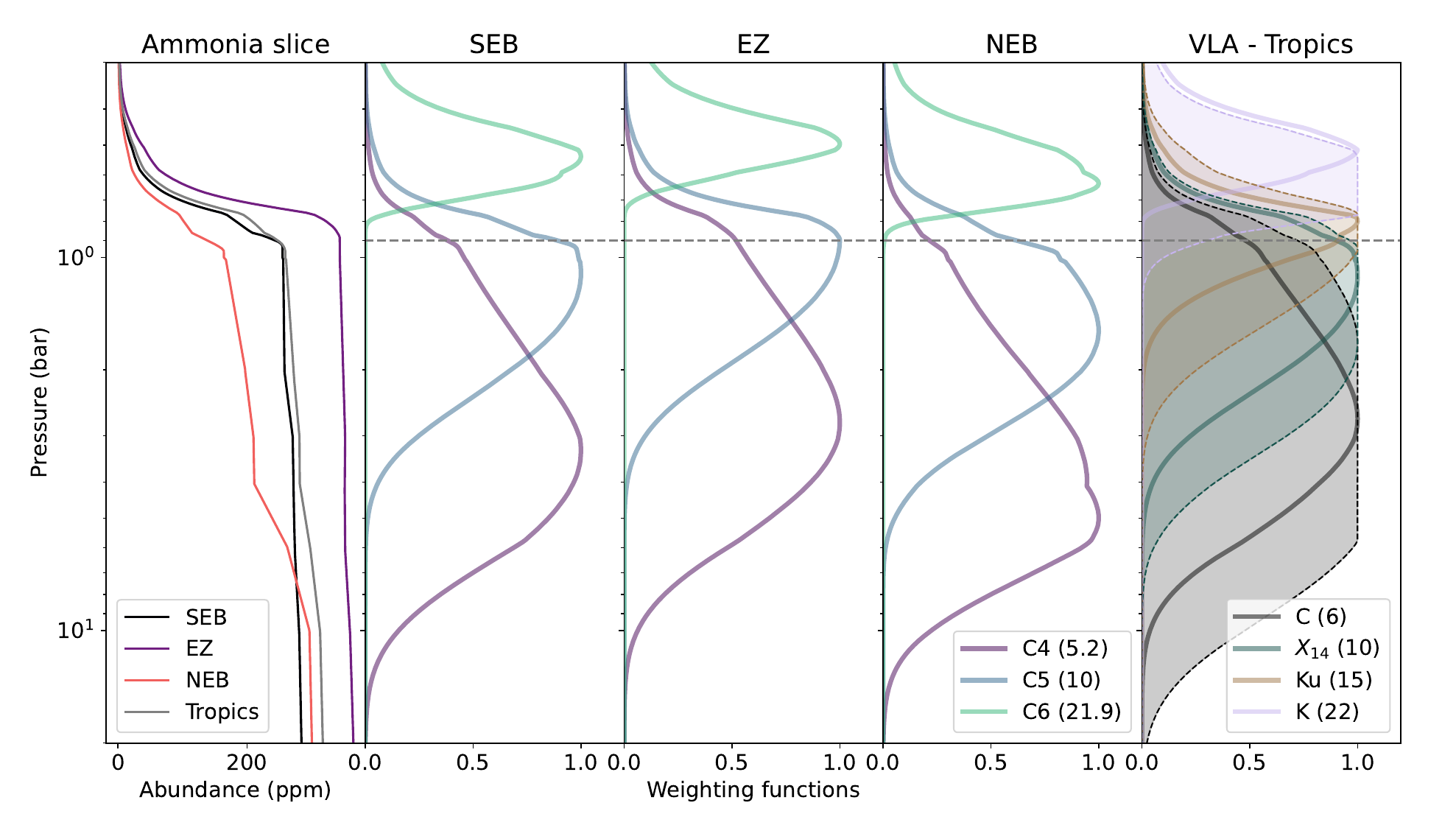}
    \caption{he leftmost panel shows the mean vertical ammonia profile for each latitude region, respectively, as indicated in the legend. The weighting functions in the rightmost four panels are based on the averaged ammonia abundance distributions for the given latitude regions from \citet{Moeckel2023}. The central three panels show the weighting functions for MWR C4, C5, and C6. The rightmost panel shows the weighting function for the VLA frequency bands used, based upon the average tropical distribution between -20 and 17{\textdegree} latitude. The shading indicates the impact of the wider VLA frequencies.}
    \label{fig:Weighting}
\end{figure}

In order to understand the depths from which we receive atmospheric information, we compute the weighting functions based on the mean vertical ammonia profile for the NEB, EZ, and SEB, respectively, coupled to a dry adiabat from \citet{Moeckel2023}. As shown in Figure \ref{fig:Weighting}, observations at 5 GHz span the cloud formation altitudes of Jupiter's atmosphere. At 10 and 15 GHz, we sample the upper few bars of the atmosphere. At 22 GHz, measurements probe within and above the ammonia ice cloud. Lower ammonia abundances in the NEB reduce the atmosphere's opacity, and the NEB measurements probe the deepest in the atmosphere. The rightmost panel shows the VLA weighting functions (the solid line is the center frequency, and the shading represents bandwidth) close to the MWR center frequencies; see Table \ref{tab:VLA}. The wide bandwidth of the VLA will widen the weighting functions, increasing the pressure range over which the signal is integrated.

\subsection{Resolution} \label{subsec:resolution}
For the interferometric measurements with the VLA, the resolution on the sky is diffraction limited, with high frequency and wider antenna spacing improving the resolution. We converted the resolution on the sky to km on the planet by multiplying the length of the beam's major axis of the VLA beam by the distance from the VLA to Jupiter at the time of observation, as given by the JPL Horizons System\footnote{\url{https://ssd.jpl.nasa.gov/horizons/}}. The beam is elliptical, so we use the major axis of the beamwidth to be conservative in our resolution calculation; beamwidth values and resolution are given in Table \ref{tab:VLA}. These resolutions are for measurements in the EZ, where the resolution is best, but the resolution only increases by 6\% at most for the northern-most and southern-most latitudes. \par

MWR observations have a variable resolution as the spacecraft's orbital geometry changes while it travels through each PJ. The values are set by the beam size and angle, the number of overlapping beams in a given region, and the distance of Juno from the ``surface". Due to the multiple variables involved, no one formula can be used to calculate MWR's best resolution. The smaller limit for the best resolution range, given in Table~\ref{tab:MWR_res}, was estimated using C5 measurements, which correspond well to visible observations. Thus, the smallest complete cloud feature seen simultaneously in C5 and concurrent, visible observations can give the lower bound on the resolution. Juno's closest distance to Jupiter multiplied by the standard beam gives the upper bound in each latitude region. \par

C4 and C5 have approximately the same beamwidth of 12{\textdegree} and C6 has a 10.8{\textdegree} beamwidth \citep{Janssen2017}. The resolution of C6 will be approximately 11\% better than C4 and C5 due to the smaller beamwidth. Each PJ has a different trajectory and thus its own resolution, but this difference is relatively small. Approximate resolutions for MWR are shown in Table \ref{tab:MWR_res}. We find that the observations in the NEB have the highest resolution, where MWR's closest approach to Jupiter occurred; the EZ and SEB have a lower resolution because these measurements were taken at larger distances. Nevertheless, we expect the deconvolved MWR observations to have a higher resolution than all the VLA observations under consideration in this work. 

\begin{table}[h!]

\caption{Juno MWR parameters. \label{tab:MWR_res}}
\centering
\begin{tabular}{ ccccccc } 
    \hline
    \hline
    Channel & Center freq & Beamwidth& Bandwidth  & NEB              & EZ                   & SEB \\ 
            & (GHz)       & (deg)    & (MHz)  & Best resolution (km) & Best resolution (km) & Best resolution (km) \\
    \hline
    C4      & 5.2         & 12.0     & 169  & 400 - 700              & 600 - 750            & 700  - 950 \\
    C5      & 10.0        & 12.0     & 325  & 400 - 700              & 600 - 750            & 700  - 950 \\
    C6      & 21.9        & 10.8     & 770  & 350 - 600              & 550 - 650            & 600 - 850 \\
    \hline
\end{tabular}
\break
\break
\raggedright
Frequency channel, center frequency, beamwidth, bandwidth, and estimated best resolution range for the C4 (5 GHz), C5 (10 GHz), and C6 (22 GHz) MWR measurements. The smaller bound on the resolution is based on the smallest complete cloud feature in C5 and visible observations, and the larger bound is based on the beamwidth and closest distance to Jupiter in each latitude region. The resolution values are best in the NEB. Center frequency, beamwidth, and bandwidth values are from \citet{Janssen2017}. 
\end{table}

\section{Results and Discussion} \label{sec:results}
\subsection{Histograms} \label{subsec:hists}
We are interested in atmospheric variability, which we characterize by considering the brightness temperature variations for a given region. We subtracted the respective mean brightness temperatures from all data sets to visualize these anomalies more clearly. The mean brightness temperatures for each region and frequency are shown in Table \ref{tab:mean_temps}. The mean temperatures show small variations between the VLA and MWR for each frequency, which are within the absolute calibration uncertainties of both instruments (1.5\% for MWR, 3\% for VLA). Multiple factors contribute to this difference, including data reduction techniques, temporal variations, spatial resolution, and the difference in bandwidth between the instruments. Removing the mean brightness temperature allows for a direct comparison of brightness temperature anomalies. \par

\begin{table}[h]
\label{tab:mean_temps}
\caption{$T_{b, \ mean}$ removed for each region and frequency.}
\hspace{30mm}
\begin{tabular}{ cccc } 
    \hline
    \hline
                     & NEB      & EZ       & SEB      \\ 
                     & $T_{b, \ mean}$ (K) & $T_{b, \ mean}$ (K) & $T_{b, \ mean}$ (K) \\
    \hline
    VLA C band 2014  & 253      & 232      & 243      \\
    MWR C4           & 258      & 232      & 245      \\
    VLA X band 2016  & 191      & 176      & 183      \\ 
    VLA X band 2014  & 196      & 186      & 186      \\
    MWR C5           & 201      & 182      & 191      \\
    VLA Ku band 2013 & 162      & 154      & 157      \\
    VLA K band 2014  & 141      & 135      & 139      \\
    MWR C6           & 145      & 138      & 142      \\
    \hline
\end{tabular}
\break
\break
\raggedright
Mean brightness temperature per latitude region and frequency removed from each set of measurements in order to compare brightness temperature distributions for Figure \ref{fig:hists}. \label{tab:mean_temps}
\end{table}

\begin{figure}[h!]
\centering
\epsscale{1.23}
\plotone{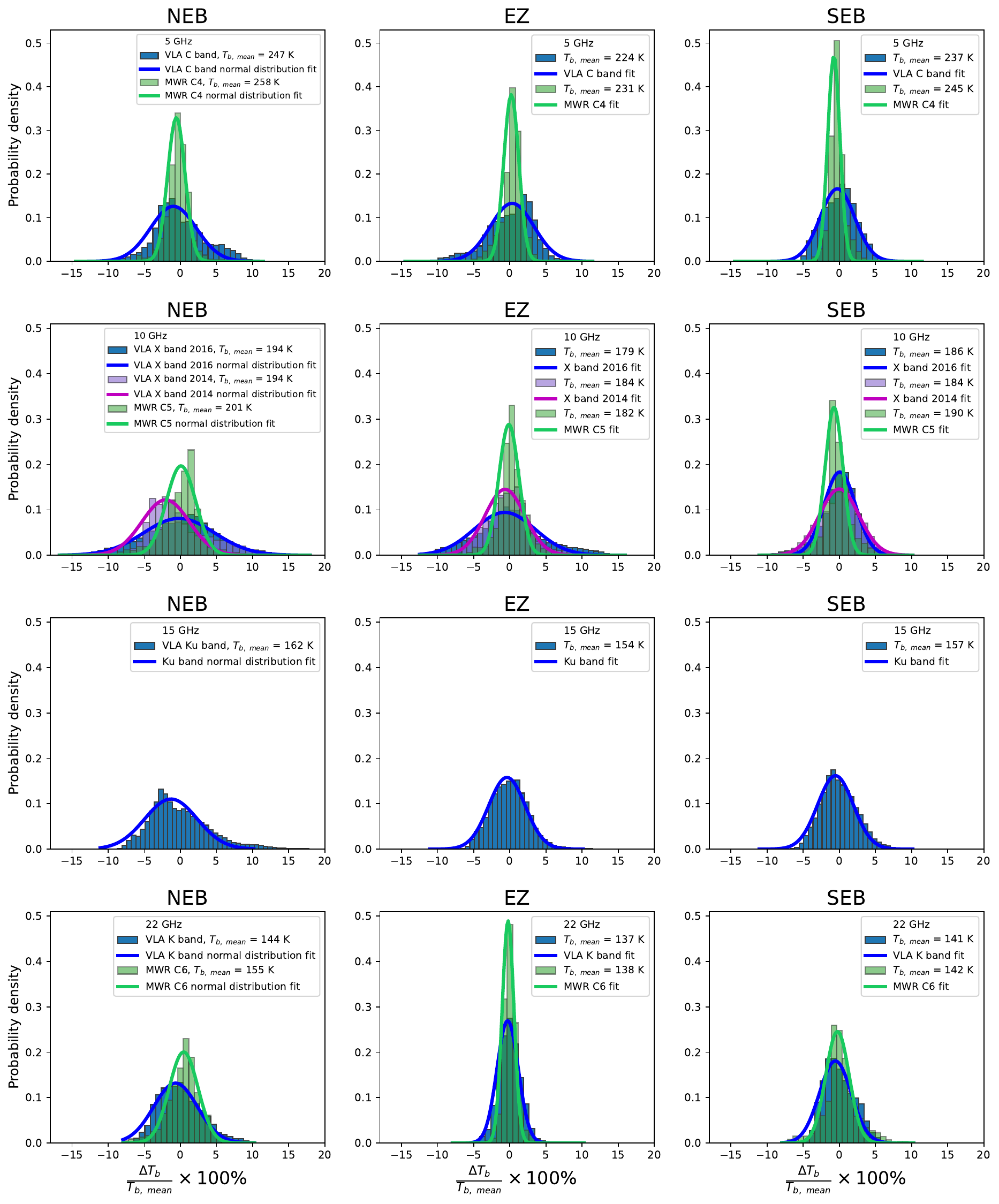}
\caption{Histograms of the NEB, EZ, and SEB comparing the variations in brightness temperature between the VLA and MWR measurements, where the brightness temperatures are plotted as percent difference from Eq.~\ref{eq:perc_diff}. Normal distributions for each are shown and the mean and standard deviation for each distribution is summarized in Table~\ref{tab:norm_distribution}.}
\label{fig:hists}
\end{figure}

Figure \ref{fig:hists} shows 12 histograms with, each column representing one of the three latitude regions under consideration, the NEB, EZ, and SEB, and each row representing one of the four frequencies under consideration. The rows contain three sets of these histograms to compare the frequencies of C4 and C band ($\sim$5 GHz), C5 and X band ($\sim$10 GHz), Ku band ($\sim$15 GHz), and C6 and K band ($\sim$22 GHz). The histograms capture information about the longitudinal brightness temperature variation within a belt or zone. To better intercompare the fluctuations between the three latitude regions, we divided each number obtained above by the mean brightness temperature in that latitude band, i.e., the values plotted in Figure \ref{fig:hists} are: 

\begin{equation}
    {\rm Percent}\ {\rm difference} = \frac{T_{b, \ obs} \ - \ T_{b, \ mean}}{T_{b, \ mean}} \times 100\%
    \label{eq:perc_diff}
\end{equation}

A normal distribution was fit to each histogram in Figure~\ref{fig:hists} to quantitatively illustrate their differences. The mean and standard deviation resulting from each fit are summarized in Table~\ref{tab:norm_distribution}. The histograms in Figure~\ref{fig:hists} are not all normally distributed, some are more symmetric than others. The histograms with more skew, such as VLA X and C band 2014, are not modeled as well by a normal distribution. The skew of the histograms also leads to most of the mean values being negative.
\par

All histograms show overall agreement between MWR and VLA. However, the distribution as seen with the VLA is usually broader than with MWR, i.e., the wings of the VLA histograms extend over a broader range, usually with a standard deviation of 2 or more greater than MWR, and/or have a higher probability density for larger deviations from $T_{b, \ mean}$. We attribute this to the fact that the VLA observes the entire globe, increasing the variety of events observed, while MWR only sees small sections of longitude. Even though longitudinal structure is visible in Figures~\ref{fig:fig1_C4} - \ref{fig:fig1_C6}, it is not enough to produce comparable anomalous tails in the MWR distributions. \par

\begin{table}[]

\caption{Mean, $\mu$, and standard deviation, $\sigma$, of the percent differences from fitting a normal distribution to each histogram in Figure~\ref{fig:hists}.}
\label{tab:norm_distribution}
\hspace{25mm}
\begin{tabular}{ccccccc}
    \hline
    \hline
            & \multicolumn{2}{c}{NEB} & \multicolumn{2}{c}{EZ} & \multicolumn{2}{c}{SEB} \\
            & $\mu$     & $\sigma$    & $\mu$     & $\sigma$   & $\mu$     & $\sigma$\\
            \hline
C band      & -0.961    & 3.16        & 0.322     & 3.00       & -0.268    & 2.4     \\
C4          & -0.554    & 1.21        & 0.201     & 1.05       & -0.759    & 0.847   \\
X band 2016 & -0.178    & 4.92        & -0.716    & 4.22       & 0.064     & 2.17    \\
X band 2014 & -2.04     & 3.27        & -0.675    & 2.75       & -0.019    & 2.73    \\
C5          & 0.086     & 2.03        & -0.134    & 1.39       & -0.739    & 1.23    \\
Ku band     & -1.38     & 3.74        & -0.489    & 2.53       & -0.613    & 2.47    \\
K band      & -0.663    & 3.03        & -0.304    & 1.48       & -0.553    & 2.21    \\
C6          & 0.459     & 1.99        & -0.244    & 0.815      & -0.261    & 1.62    \\
\hline
\end{tabular}
\end{table}
The NEB has a normal distribution with long warm and cold tails across all frequencies, and the SEB shows the smallest range of brightness temperature anomalies. The EZ has a similarly wide distribution, like the NEB. However, in the EZ we find more variability between the different frequencies. For example, the distribution in the EZ has a warm tail at 10 GHz and a cool tail at 5 GHz. This variation with depth is not seen in the NEB or SEB. Though the atmosphere is constantly changing, there is no evidence of significant changes to composition or structure over the period of time under consideration, 2013 to 2018.\par

We acknowledge that changes in resolution affect the width of the distribution, with higher-resolution observations resolving more fine-scale extremes. In the NEB, we see the widest spread of brightness temperature anomalies in all MWR channels and regions considered, coinciding with the region where the MWR resolution is highest due to the orbital geometry. Resolution effects could partially explain the broader distribution in the NEB compared to the SEB for MWR measurements. However,the VLA measurements show a similar difference between the regions despite the resolution being hemispherically symmetric. Hence, we interpret these differences to be of a dynamic nature and not a resolution effect. The widest distribution of all data sets is in the NEB VLA X band 2016 measurements, where the resolution approaches the best resolution measured by MWR, yet MWR does not see as broad a distribution due to its limited view in longitude. \par

The measurements for VLA X band 2014 have more extreme events than those for K band 2014 despite having a similar resolution, implying more dynamic activity below the ammonia ice cloud than above. The C band 2014 and X band 2014 observations have distributions of similar width, despite the C band measurements having better resolution, which implies that the atmosphere as measured by X band is at least as variable, likely more variable, than C band.

\subsection{Interpretation of Distributions}
The different frequencies used in our analysis inform us about the altitude at which the dynamics occur in the atmosphere. We find the widest distribution of brightness temperature anomaly at 10 and 15 GHz. We interpret this to be primarily caused by the condensation and re-evaporation of ammonia, the tell-tale sign of weather activity on Jupiter. At 10 and 15 GHz, probing just below the cloud layer, both distributions display wider spreads of brightness temperature anomaly than the 22 GHz measurements. The smaller variation in brightness temperature at 22 GHz might be expected; the ammonia abundance approximately follows the saturated vapor curve \citep{dePater2019_VLA}. Higher ammonia abundances compensate for increases in physical temperature, so the brightness temperature stays approximately constant. \par

Due to the wider spread of brightness temperature anomaly in the NEB, as evidenced by the increased standard deviations in Table~\ref{tab:norm_distribution}, we interpret the NEB as the most dynamic of the three regions. At the boundary of the NEB and EZ, the equatorially trapped Rossby wave creates alternating regions of ammonia depletion and abundance \citep{showman_2000}, also referred to as ammonia hot spots and plumes, respectively. The hot spots in the NEB, coupled with the plumes in the EZ, could help explain the long tails in their respective distributions.
This requires a planetary-scale wave pattern to produce these long tails. Juno has skimmed a few hotspots but does not capture the full spatial extent \citep{Fletcher2020}.\par

The SEB is less warm on average than the NEB, as shown in Table \ref{tab:mean_temps}, and likewise has a narrower distribution than the other two regions, indicating that despite similar solar heating conditions between the NEB and SEB, different processes must be driving both regions. This is surprising, given that they are both belts near the equator. However, the presence of the Great Red Spot may affect the SEB. The wide distribution in the EZ indicates active dynamics obscured by the visible cloud tops.\par

\subsection{Locations of High Anomaly}

\begin{figure}[h!]
\centering
\epsscale{1.2}
\plotone{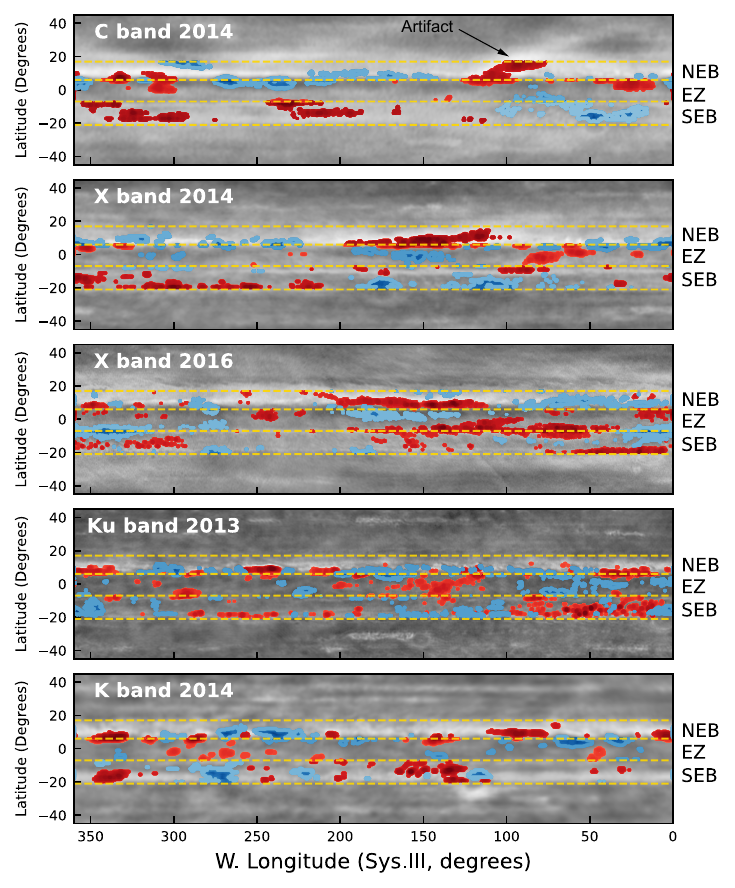}
\caption{VLA maps showing the locations of the strongest positive and negative brightness temperature anomalies within the NEB, EZ, and SEB. The grayscale VLA maps are the same as Figure~\ref{fig:VLA maps}. Locations of warm brightness temperature anomaly are shown in red and represent the top 5\% of anomalies. The negative brightness temperature anomalies are shown in blue and represent the bottom 5\% of values. Darker colors for both indicate more extreme brightness temperatures. Unlike Figure~\ref{fig:hists}, the anomalies were zonally detrended, as opposed to one $T_{b, \ mean}$ per zone or belt as we did for Figure \ref{fig:hists}.}
\label{fig:extremes}
\end{figure}

We are interested in the location and scale of brightness temperature anomalies to help identify processes that can enhance or deplete the atmosphere in trace gases. Figure~\ref{fig:extremes} visualizes the locations and scale of the strongest anomalies in the data by highlighting the top 5\%, in red, and bottom 5\%, in blue, of the brightness temperature values in the VLA measurements. To describe the extreme brightness temperature trends in Figure~\ref{fig:extremes}, it is helpful to define a longitude scale. We consider a feature that extends over 1{\textdegree} of longitude to be small-scale (i.e., $\lesssim$ the Rossby deformation radius of $\sim$1500 km \citep{depater2010vortex,young2017}), 5 – 10{\textdegree} to be intermediate-scale, and more than 30{\textdegree} to be large-scale. Unlike Figure~\ref{fig:hists}, we de-trended the observations by removing the zonal mean to focus on the longitudinal variability for a given latitude as opposed to one $T_{b, \ mean}$ per latitude region, where we focused on the climatological signal of the zones and belts. 

In general, there is a wide range of anomaly scales spatially, but we find interesting differences between their morphology. The warm anomalies, which we interpret as areas of ammonia depletion, appear in two modes: large-scale provinces and small-scale features. The cold anomalies, which point towards an enhancement in ammonia, fall more often on the intermediate scale. 

The different frequencies show distinct trends in Figure~\ref{fig:extremes}. In C band, mostly large-scale signals are visible despite the higher resolution of C band than the X band 2014 data. Measurements in X and Ku bands probe the weather layer and show the widest range of anomaly scale, displaying many more small-scale anomalies than C or K bands. A similar trend to the C band map is seen in K band, with K band primarily showing intermediate- or large-scale anomalies, implying that only the most extreme events reach that height in the atmosphere.

In the NEB, the Ku band and both X band maps show more large-scale anomalies than the SEB. We interpret the large-scale features as part of the equatorial trapped Rossby wave that creates the 5-micron hotspots \citep{allison_1990,Ortiz1998}, and the cold plumes in the equatorial region \citep{dePater2016, fletcher2016}. The NEB does not show many intermediate-scale anomalies in either X band map. In Ku band, the NEB has anomalies primarily in its southern latitude boundary. The EZ has fewer anomalies than the NEB or SEB. The average scale of the anomalies in the EZ are generally similar to that of the NEB and SEB.

The SEB has more small-scale anomalies in Ku band and both X band maps. Like the NEB, the SEB does not show many intermediate-scale anomalies in either X band map. One area of interest in the SEB is the X band 2016 data from 300 - 360{\textdegree} longitude, which shows more small-scale activity than anywhere else in the four maps. This could indicate increased storm activity in the SEB, which fits with the increased amount of lightning observed by \citet{BrownShannon2018Plsa} in the SEB compared with the NEB. Another reason this region is of interest is that the increased storm activity is an indicator of large-scale outbreaks \citep{Fletcher2017}. Indeed, a month after the X band 2016 observations, \citet{dePater2019_alma} reported on such an outbreak in the SEB.

When looking at the morphology of the anomalies, we find them predominantly elongated in the longitudinal direction and seemingly close to the interfaces between zones and belts. We interpret that the processes that create localized anomalies interact with the jets, while mixing in the meridional direction is inhibited by the atmosphere \citep{Read2006}. This requires that these events remain active long enough to propagate around the planet. Not all localized events propagate in longitude, as seen with the collection of small, warm anomalies in the X band 2016 data in the SEB. Some plumes and hotspots remain visible and confined longitudinally. These small-scale storms may remain localized because they are too short-lived to propagate longitudinally. The intermediate-scale structure, i.e. plumes and hot spots, are observed to have a velocity that allows them to remain somewhat spatially confined in longitude \citep{Ortiz1998}.  The area of warm anomalies is greater than that of cold anomalies, which could mean that ammonia depletion has a longer timescale than ammonia enhancement. 

Considering all four frequencies in conjunction, we see that the majority of variability is seen in the weather layer as probed by the X and Ku band observations. This implies that small-scale anomalies, which we interpret as localized storms, are mostly affecting pressures around ammonia cloud formation and only the most violent processes can perturb the atmosphere both below (C band) and above (K band) the weather layer.

\section{Conclusions} \label{sec:conclusions}
We used a combination of MWR and VLA brightness temperature anomalies to trace dynamics in Jupiter's NEB, EZ, and SEB. We base our analysis on brightness temperature maps for four distinct frequency bands (5 GHz, 10 GHz, 15 GHz, and 22 GHz). The VLA maps are global and integrated over a Jovian rotation. In contrast, the MWR maps cover approximately a quarter of the planet based on nine individual PJs spanning approximately two years of observation. Although the MWR data only sample a narrow swath in longitude during each PJ, we used the entire longitudinal extent of each PJ to get maximum longitudinal coverage. We note that this is the first time such maps are shown from the MWR data. We first separated the maps into NEB, EZ, and SEB latitude regions for our analysis. We then removed the mean brightness temperature for the three regions and four frequencies to avoid calibration bias between the two instruments. We computed the distribution in Figure~\ref{fig:hists} that displays the range of brightness temperature anomaly and a normal distribution fit for each frequency and region. The mean and standard deviations for each normal distribution fit are summarized in Table~\ref{tab:norm_distribution}. \par

The frequencies used correspond to different depths in Jupiter's atmosphere (Figure \ref{fig:Weighting}), and the signals we use originate from within and above the ammonia condensation pressure (22 GHz), just below the ammonia condensation pressure (10 and 15 GHz), and the region between the water cloud and ammonia clouds (5 GHz). We find the widest distribution of anomalies centered below the ammonia ice clouds (10 and 15 GHz). The observations probing deeper into the atmosphere (5 GHz) have significant, albeit fewer anomalies, whereas we find the least amount of variability highest in the atmosphere (22 GHz). Taken together, this indicates that most ammonia or temperature variability happen in the weather layer of Jupiter, primarily between 1 - 3 bars. The wider distributions of the VLA indicate that Jupiter's atmosphere is more extreme than that sampled by Juno during its nominal mission. Despite MWR's better spatial resolution, both telescopes predominantly sample atmospheric features controlled by geostrophic balance. We expect more extreme events to occur on convectively controlled scales, i.e., at scales at or smaller than the Rossby deformation radius. The chances that Juno, with its longitude-limited view, would capture such a small feature are very small compared to the global-mapping capabilities of the VLA.

Though the histograms in Figure~\ref{fig:hists} capture the numerical distribution of brightness temperature anomalies for each region, the physical scale associated with the features causing these anomalies is not constrained. Instead, we visualized the spatial scale of anomalies in Figure~\ref{fig:extremes}, where we map the most extreme values measured by VLA and study their morphology. The different spatial scales visible in the anomalies point to different dynamics shaping the atmosphere, ranging from localized storm systems in the SEB to large-scale structures in the NEB. In our observations, we see an increased number of small-scale anomalies in the SEB, consistent with localized water storms and, in this area, a storm broke out a month after the data in this report \citep{dePater2019_alma}. We find the largest-scale brightness temperature anomalies in the NEB, which are consistent with the 5-micron hotspots set by an equatorially trapped Rossby wave. Yet the absence of small-scale anomalies in this region remains surprising. We find that most anomalies are extended in longitude, indicating that meridional mixing must be inhibited. 

In summary, distinct trends in atmospheric dynamics were observed in each of the latitude regions and depths in the atmosphere considered. We find evidence for small-scale events primarily in the weather layer (10 and 15 GHz) and larger-scale structure is seen both deeper (5 GHz) and higher (22 GHz) in the atmosphere. Our findings suggest that the NEB and SEB have different processes shaping them, despite being hemispherically symmetric. Of the three regions considered the SEB has the least dynamical imprint across all frequencies, while the NEB and EZ are significantly more variable. 

In the future, concurrent observations of Jupiter's atmosphere in optical and infrared wavelengths can help to couple cloud to sub-cloud dynamics observed with radio telescopes. Further modeling work is necessary to understand the origin and lifetime of the observed anomalies.

\begin{acknowledgments}
We thank Huazhi Ge for help with interpretation of the results and the referees for improving the rigor of the manuscript. The VLA data used in this report,
associated with project code 13B-064 (2013–2014) and 16B-
048 (2016), are available from the NRAO Science Data
Archive at https://archive.nrao.edu/archive/advquery.jsp. The
National Radio Astronomy Observatory is a facility of the
National Science Foundation operated under cooperative
agreement by Associated Universities, Inc.
This research was supported by NASA’s Solar System Observations
(SSO) award 80NSSC18K1003 to the University of
California, Berkeley.
Joanna Hardesty was participating in the Undergraduate Research Apprentice
Program (URAP) at the University of California, Berkeley while conducting part of this research.
\end{acknowledgments}

\appendix
\section{VLA map correction}
\begin{figure}[h!]
\centering
\epsscale{1.1}
\plottwo{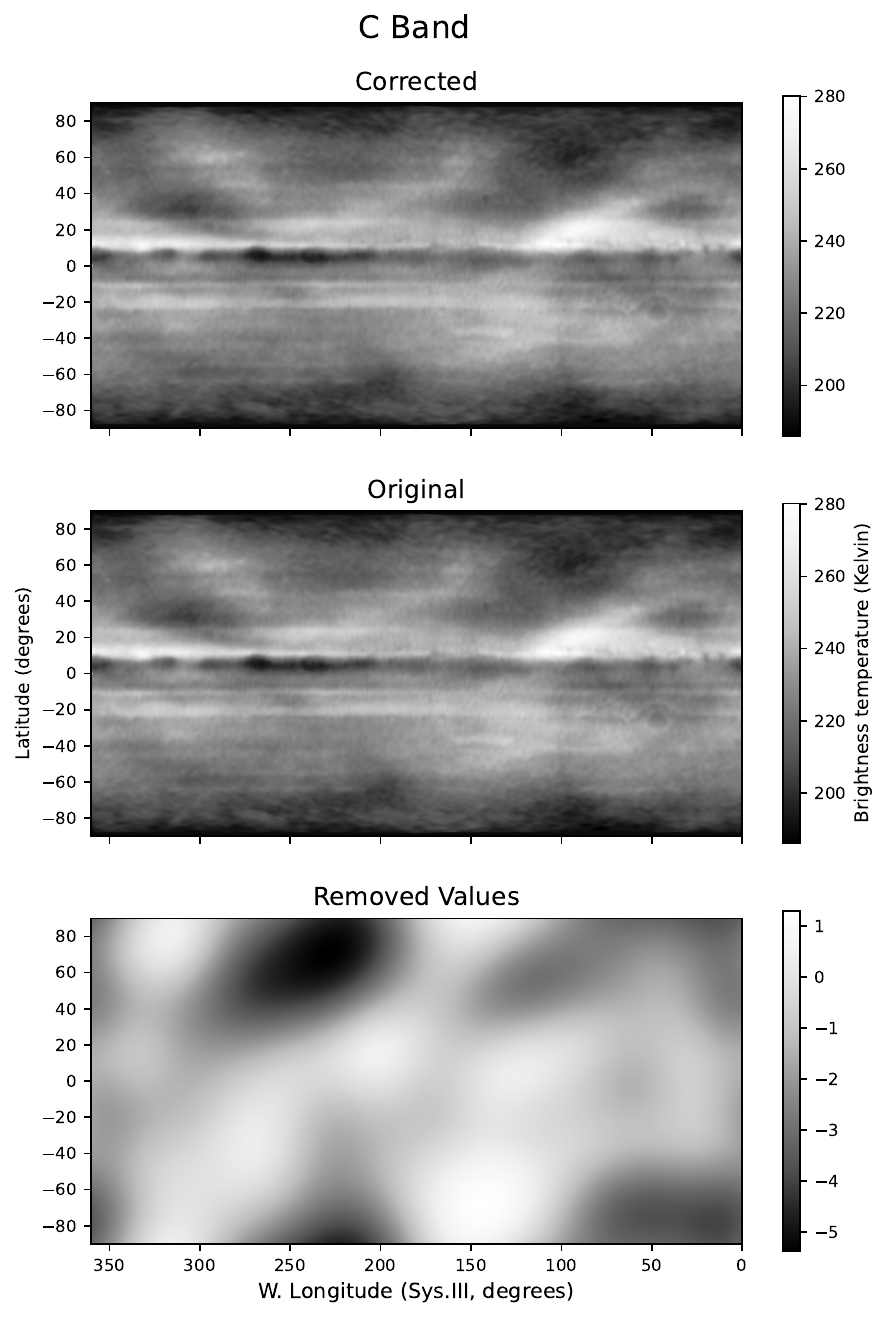}{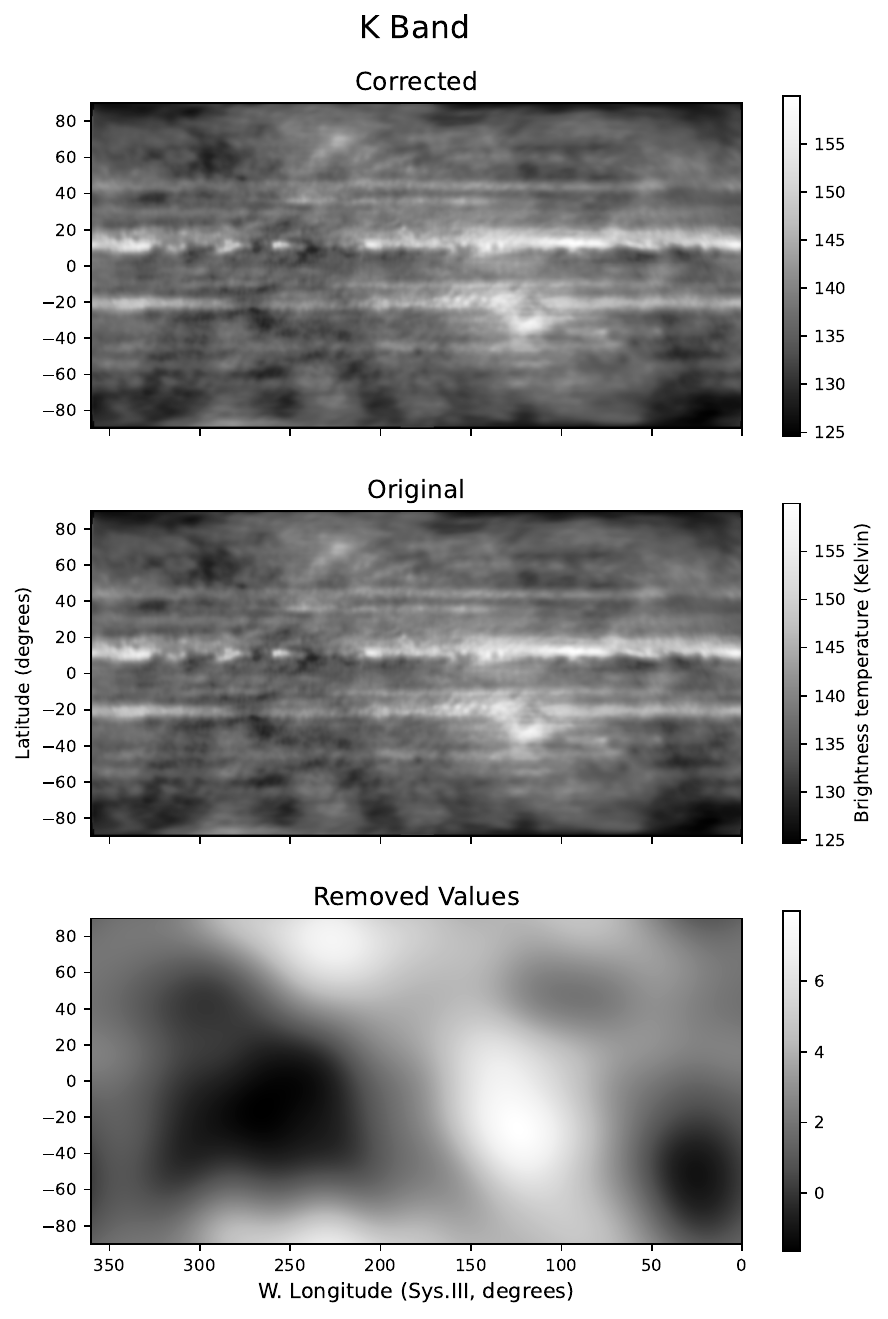}
\caption{VLA C band 2014 map showing artifact correction with an 80-pixel Gaussian filter, left, and K band 2014 map showing correction with a 40-pixel Gaussian filter, right.}
\label{fig:c-kband-correction}
\end{figure}

\begin{figure}[h!]
\centering
\epsscale{1.1}
\plottwo{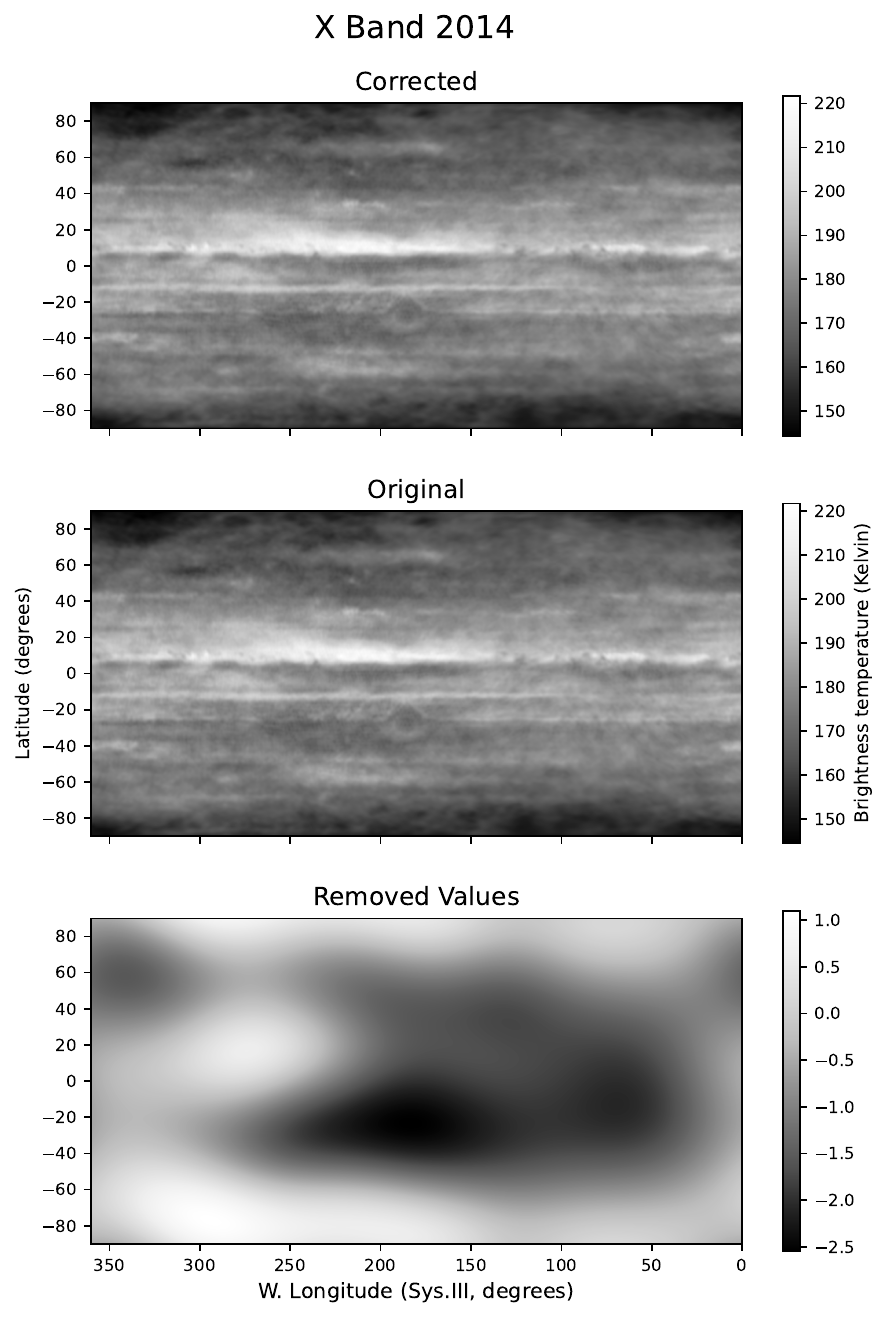}{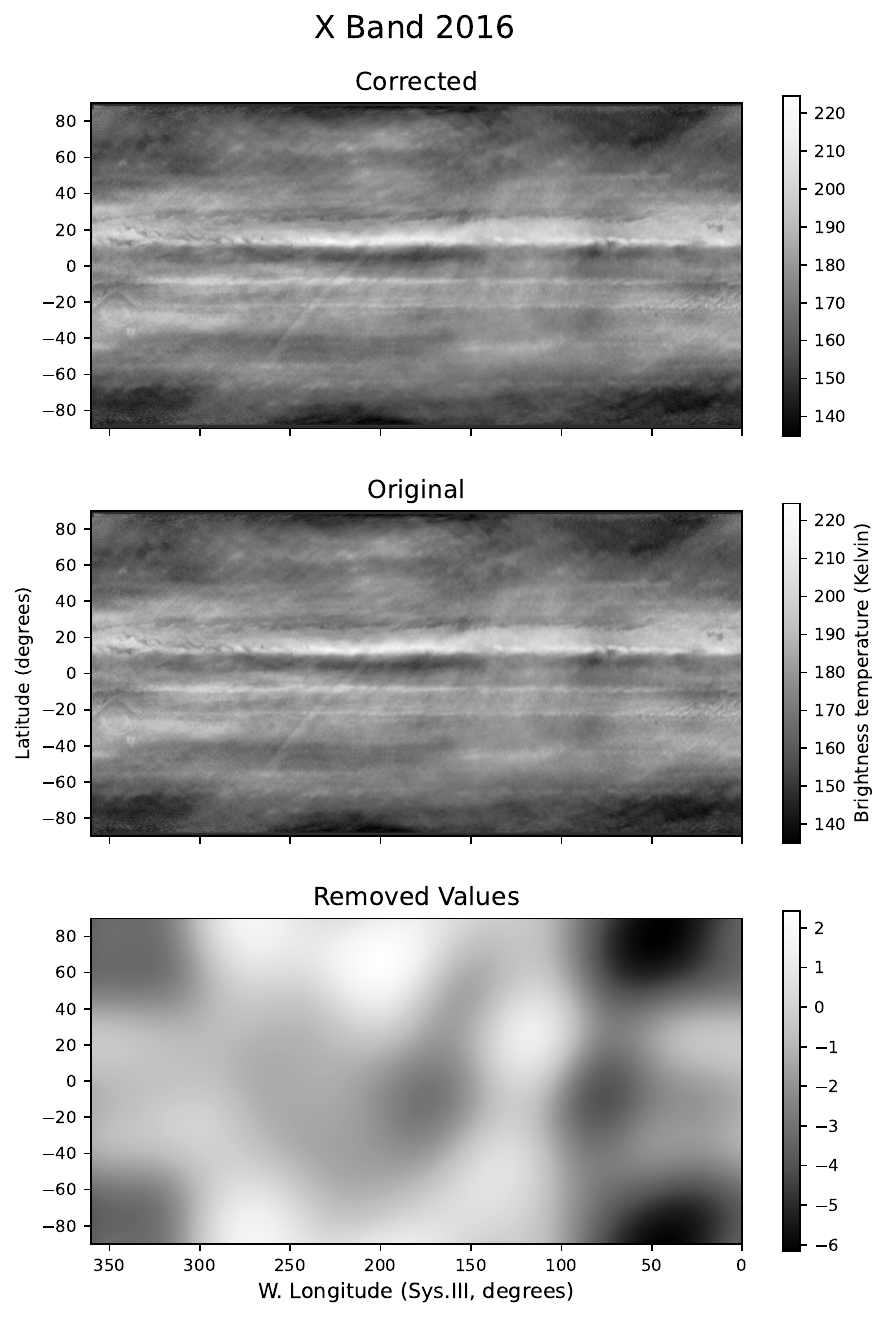}
\caption{VLA X band 2014 map showing artifact correction with an 80-pixel Gaussian filter, left, and X band 2016 map showing correction with a 120-pixel Gaussian filter, right.}
\label{fig:xband-correction}
\end{figure}

\begin{figure}[h!]
\centering
\epsscale{0.6}
\plotone{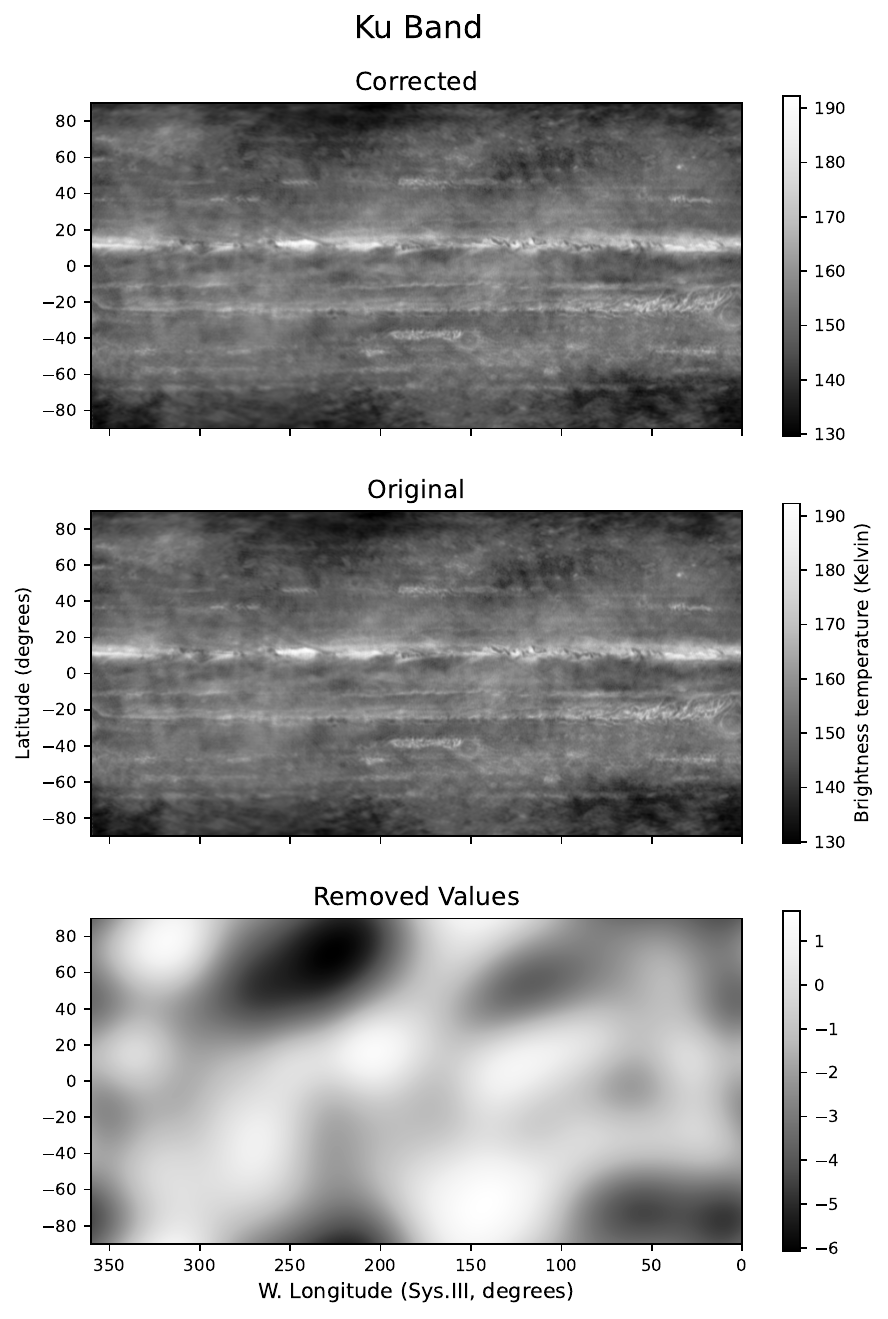}
\caption{VLA Ku band 2013 map showing artifact correction with a 65-pixel Gaussian filter.}
\label{fig:kuband-correction}
\end{figure}

In order to reduce bias in the VLA maps, primarily due to synchrotron radiation and undersampling of the \textit{uv}-plane for the interferometric inversion, we corrected the VLA maps to remove large-scale artifacts. The synchrotron radiation will always appear as a warm signal, usually on the order of a few degrees Kelvin. Artifacts will always be large-scale signal, multiple degrees in longitude or greater, therefore small-scale features in the maps can be considered signal. To mitigate the artifacts in the data, we applied a Gaussian, or high-pass, filter to the maps and removed that signal. The kernel size of the filter can be adjusted to only remove large-scale structures, which we know to be outside of the maximum resolution limit of our radio measurements. This method is conservative and has the potential to remove actual signal from our maps, but will not add false signal or contribute to additional artifacts. Removed structure and resultant maps can be found in Figures \ref{fig:c-kband-correction}, \ref{fig:xband-correction}, and \ref{fig:kuband-correction}. We tried multiple different filters to see which removed the most artifacts without greatly reducing the signal. While removing the signal in this filter did help to reduce artifacts, our results are robust to the chosen filter size. We tried many different filters without fundamentally changing the results in this paper.

\bibliography{main}{}
\bibliographystyle{aasjournal}

\end{document}